\documentclass[runningheads]{llncs}

\usepackage{wrapfig}
\usepackage{graphicx}
\usepackage{multirow}
\usepackage{subfigure}
\begin{document}
\title{ViT3D Alignment of LLaMA3: 3D Medical Image Report Generation}
\author{Siyou Li\inst{1}\orcidID{0009-0007-6840-5791} 
\and
Beining Xu\inst{1}\orcidID{0009-0000-3342-1782} \and
Yihao Luo\inst{2}\orcidID{0009-0002-1169-2930} \and Dong Nie\inst{3}\orcidID{0000-0003-0385-8988} \and Le Zhang\inst{4}\orcidID{0000-0002-3848-0017}
}
\authorrunning{Siyou Li et al.}
\institute{School of Electronic Engineering and Computer Science, \\Queen Mary University of London, London, UK\\
\email{\{s.li, b.xu\}@se23.qmul.ac.uk}\\ 
\and
Department of Bioengineering, Imperial College London, London, UK\\
\email{y.luo23@imperial.ac.uk}
\and
Meta Inc. US\\
\email{dongnie@cs.unc.edu}
\and
School of Engineering, College of Engineering and Physical Sciences, \\University of Birmingham, Birmingham, UK\\ \email{l.zhang.16@bham.ac.uk}
}
\maketitle              % typeset the header of the contribution
\begin{abstract}

Automatic medical report generation (MRG), which aims to produce detailed text reports from medical images, has emerged as a critical task in this domain. MRG systems can enhance radiological workflows by reducing the time and effort required for report writing, thereby improving diagnostic efficiency. In this work, we present a novel approach for automatic MRG utilizing a multimodal large language model.  Specifically, we employed the 3D Vision Transformer (ViT3D) image encoder introduced from M3D-CLIP to process 3D scans and use the Asclepius-Llama3-8B as the language model to generate the text reports by auto-regressive decoding. The experiment shows our model achieved an average Green score of 0.3 on the MRG task validation set and an average accuracy of 0.61 on the visual question answering (VQA) task validation set, outperforming the baseline model. Our approach demonstrates the effectiveness of the ViT3D alignment of LLaMA3 for automatic MRG and VQA tasks by tuning the model on a small dataset.
\keywords{Medical Report Generation  \and Computed Tomography (CT) Imaging \and Natural Language Processing (NLP).}
\end{abstract}
\section{Introduction}

\begin{figure}[h!]
    \centering
    \includegraphics[width=\textwidth]{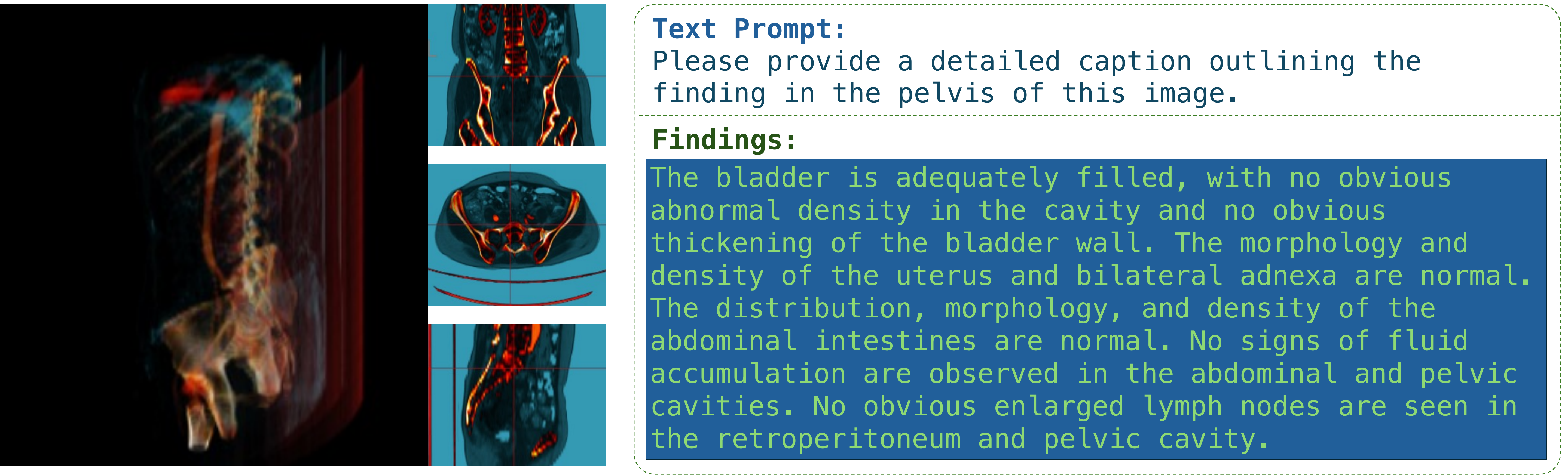}
    \caption{Example of Medical Report Generation (MRG) results from our model.}
    \label{fig:example}
\end{figure}

Computed tomography (CT) scans are sophisticated imaging technologies that use rotational X-ray beams to capture high-resolution cross-sectional images of the body. These images can be reconstructed into three-dimensional (3D) visualizations, allowing for in-depth examination of internal structures such as organs, tissues, and bones. This capability is crucial for diagnosing a broad spectrum of pathological conditions, including cancers, cardiovascular disorders, traumatic injuries, and infections. The widespread use of CT scans across various medical specialties underscores their essential role in modern healthcare, enabling clinicians to obtain precise diagnostic information that significantly aids in patient evaluation, treatment planning, and monitoring \cite{schockel_developments_2020}. Globally, the substantial volume of CT studies performed each year reflects the technology’s critical contribution to contemporary clinical practice.

The interpretation of CT scans is typically performed by specialist radiologists, who analyze the images to detect abnormalities and generate comprehensive written reports. These reports serve as an essential communication tool between radiologists and referring physicians or surgeons, detailing the radiologist’s findings, providing diagnostic conclusions, and often suggesting recommendations for additional investigations or therapeutic interventions. The process of report generation requires a high level of expertise and attention to detail, as accurate interpretation is critical for patient management and treatment planning \cite{zhao2023radiology}. However, the increasing volume of CT examinations has placed significant pressure on radiologists, highlighting the need for more efficient solutions.

Emerging natural language processing (NLP) and artificial intelligence (AI) technologies offer promising avenues for automating the generation and processing of radiology reports \cite{thirunavukarasu_large_2023}. Automated report generation systems have the potential to assist radiologists by streamlining the workflow, reducing reporting time, and minimizing the variability in report quality (see Fig.~\ref{fig:example}). Additionally, such systems can facilitate large-scale data extraction for clinical research, enabling more effective utilization of radiology data. The integration of AI-driven solutions into clinical practice could revolutionize the field, enhancing diagnostic accuracy, improving patient outcomes, and supporting the evolving demands of healthcare systems. As research progresses, these technologies are expected to become integral components of radiological practice, augmenting the capabilities of human experts and transforming the landscape of medical imaging.

\begin{figure}[h!]
    \centering
    \includegraphics[width=\textwidth]{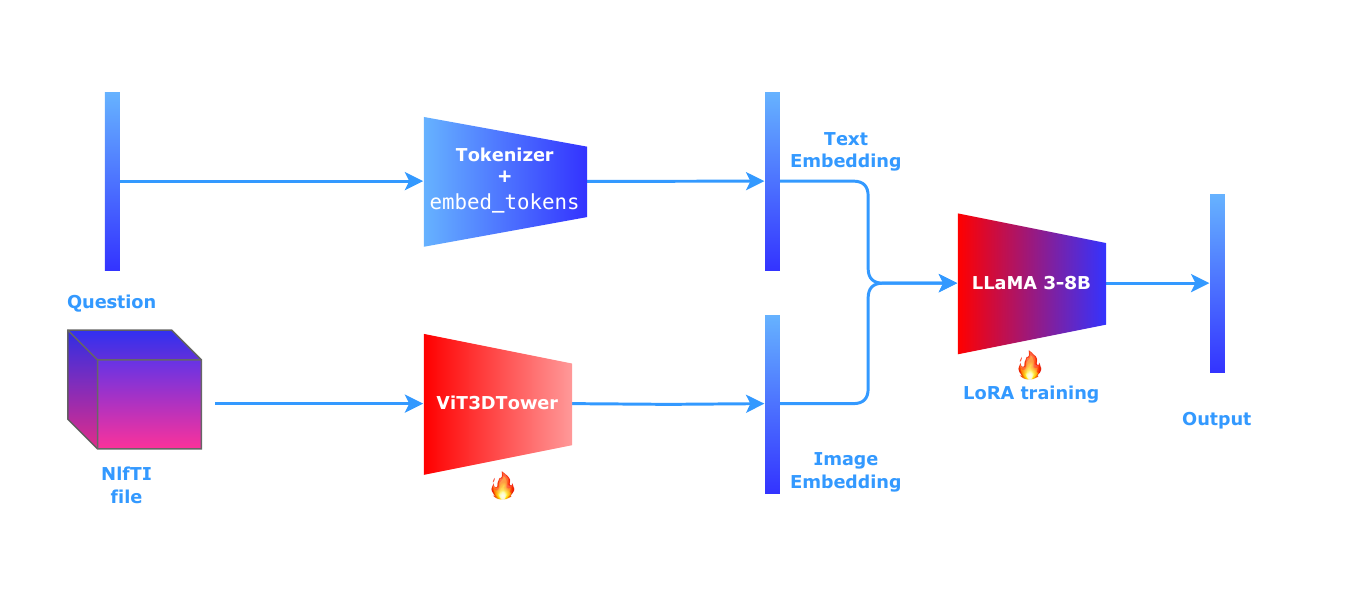}
    \caption{An overview of our ViT3D Alignment of LLaMA3 approach. The vision encoder is integrated to embed the raw image data into a feature map and then feed the feature map into a ViT module to turn the feature map into a sequence of embedding. The embedding are then concatenated with the prompt embedding and passed to a Large Language Model to generate the next token of pending-generated text.}
    \label{fig:pipeline}
\end{figure}

In this work, we propose a novel method for MICCAI2024 AMOS-MM Challenge \cite{ji2022amos} about automatic MRG using a multimodal large language model. Technically, we integrate the vision encoder to embed the raw image data into a feature map and then feed the feature map into a Vision Transformer (ViT) module to turn the feature map into a sequence of embeddings. The embeddings are then concatenated with the prompt embeddings and passed to a Large Language Model to generate the next token of pending-generated text. We fine-tune the model on the given dataset to adapt it to the MRG task. More importantly, we aligned 3D NIfTI images with a large language model using a small amount of data based on a pre-trained model. We used the ViT3D image encoder from M3D-CLIP and the Asclepius-Llama3-8B language model with a Spatial Pooling layer (see Fig.~\ref{fig:pipeline}). During training, we performed full fine-tuning for ViT3D and utilized LoRa fine-tuning for Asclepius-Llama3-8B. Our final average Green score on the MRG task validation set is 0.3, and our average accuracy on the VQA task validation set is 0.61.

\section{Method}\vspace{-1em}

\subsection{Model Architecture}
The architecture of our model is shown in Figure \ref{fig:pipeline}. Our model consists of a Vision Encoder to process the raw image data into embeddings, and a Large Language Model to generate the relevant reports corresponding to the input images, according to text prompts. Our language module is a transformer-based model that integrates the information of the current sequence of tokens and image embeddings through the attention mechanism \cite{vaswani2023attentionneed} and then predicts the logit of the next token. The Vision Encoder is based on a 3D Vision Transformer, which sequentially processes the 3D image data into embeddings. The Vision Encoder and the Large Language Model are pre-trained on a large dataset, and fine-tuned on the given dataset to adapt them to the MRG and VQA task. Here, we used Asclepius-Llama3-8B \cite{kweon2023publicly} as the language processing module and adopt the 3D ViT module from M3D-LaMed \cite{bai2024m3d} as the vision processing module. The whole text generation is achieved by the auto-regressive decoding process, where the model generates the next token of the text output based on the previous tokens, fomulating as
\begin{equation}
    P(Y| C_{pro}, C_{img}) = \prod_{t=1}^{T} P(y_t| C_{pro}, C_{img}, Y_{1:t-1}),
\end{equation}
where $Y$ is the generated text, $C_{pro}$ is the prompt embedding, $C_{img}$ is the image embedding, and $y_t$ is the $t$-th token of the generated text. The tokenizor and codebook are inherited from the LLaMA 3-8B model. ViT module will be fully trained while the LLaMA module will be fine-tuned via Low-ranking Adaptation \cite{hu2021loralowrankadaptationlarge}.

We adopt Cross-Entropy (CE) \cite{zhang2018generalized} loss as the training objective, which is defined as
\begin{equation}
    \mathcal{L} = -\sum_{t=1}^{T} \log P(y_t| C_{pro}, C_{img}, Y_{1:t-1}),
\end{equation}
where the loss minimizes the negative log-likelihood of the ground-truth text given the image embedding and the generated text.

% \subsection{Vision Encoder}
% M3D-LaMed utilizes a 3D Vision Transformer\cite{dosovitskiy2021imageworth16x16words}, which is a transformer-based model that can process 3D image data. The Vision Transformer consists of a series of transformer blocks, each of which contains a multi-head self-attention mechanism and a feed-forward neural network. The Vision Transformer processes the input image data into embeddings, which are then passed to the Large Language Model.

\subsection{Dataset}
\textbf{AMOS-MM 2024 Dataset.} Our training is mainly based on the dataset provided by the AMOS-MM 2024 challenge. The dataset consists of 2,088 medical images of the chest, abdomen, and pelvis, as well as corresponding text reports. The medical images are CT scans with spacial resolution from 256x256 to 1024x1024 and the slice thickness from 1mm to 5mm.  The text reports are manually annotated for two tasks: Medical Report Generation (MRG) and Visual Question Answering (VQA). The former and contains detailed descriptions of the findings in the images and the latter contains multiple-choice questions. The dataset is split into a training set of 1,288 images and a validation set of 400 images. We use the training set to train our model, and the validation set to evaluate the performance of our model. All the images data and reports are collected from Longgang District Central  Hospital, Shenzhen, China.
\\

\noindent 
\textbf{External Dataset.} For more robust training, better generalization and higher scalability, we involve a large-scale external dataset, CT-RATE \cite{hamamci2024foundation}, which consists of 50,188 CT images of 21,340 patients and corresponding text reports. The dataset includes images for multi-organ diseases mainly in chest. The scanning resolution and slice numbers range from 256x256 to 1024x1024 and 46 to 1277, respectively. All the images and reports are collected from Istanbul Medipol University Mega Hospital, Istanbul, Turkey. For better coupling with the given dataset, we use Named Entity Recognition (NER) \cite{li2020survey} to extract the organs from the text reports and classify the sentences into "chest", "abdomen" and "pelvis" based on the organs mentioned in the sentences.

\section{Experiments}
\subsection{Implementation}
We used the Hugging Face \texttt{Transformers} library to load the pre-trained Vision Transformer and Large Language Model. We used the \texttt{AdamW} optimizer with a learning rate of $1e-5$ to train our model. There are $0.5B$ trainable parameters in vision encoder, $59M$ in multimodal projector, and $1.1B$ in LoRA. We trained our model on 6 Nvidia A800 GPUs with a batch size of 6. We trained our model for 8 epochs on the given dataset and the external dataset.

\subsection{Results}

\begin{figure}
    \centering
    \includegraphics[width=1\textwidth]{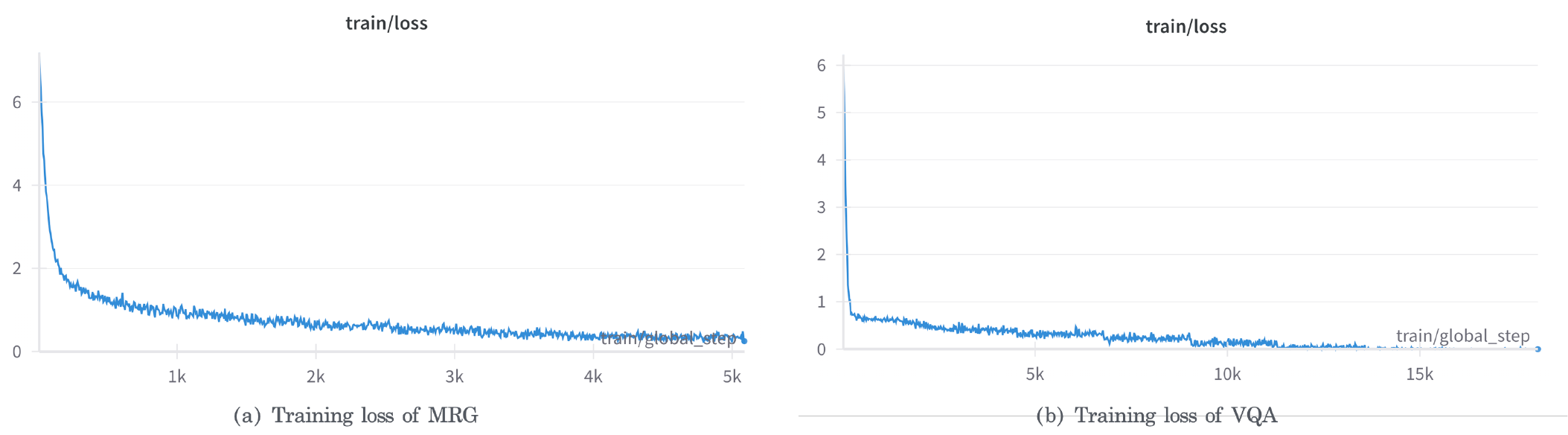}
    \caption{The curve of training loss for MRG (left) and VQA (right).}
    \label{fig:loss}
\end{figure}

The training losses on two downstream tasks are shown in Fig.~\ref{fig:loss}. The training loss of MRG converges to 0.36 after 4k steps, and the loss of VQA converges to 0.03 after 10k steps.

The evaluation results presented in Table 1 compare the performance of different models on the MRG and VQA tasks using the given validation set. For the MRG task, we employ the Green score \cite{ostmeier2024greengenerativeradiologyreport} as the evaluation metric, which quantifies the semantic similarity between generated reports and ground-truth texts based on a pre-trained language model. For the VQA task, accuracy is used as the evaluation metric, representing the percentage of correct answers to the questions posed.

\setlength{\tabcolsep}{4mm}
\renewcommand{\arraystretch}{1.1}
    
% \usepackage{multirow}
\begin{table}[]
\centering
\caption{Task 1 (MRG) and Task 2 (VQA) results on the given validation set.}
\label{tab:results}
\begin{tabular}{|l|c|cccc|}
\hline
 &
  \multirow{2}{*}{\begin{tabular}[c]{@{}c@{}}VQA\\ (Accuracy)\end{tabular}} &
  \multicolumn{4}{c|}{\begin{tabular}[c]{@{}c@{}}MRG (Green Score)\\ \end{tabular}} \\ \cline{3-6} 
          &      & \multicolumn{1}{c|}{Chest}         & \multicolumn{1}{c|}{Abdomen} & \multicolumn{1}{c|}{Pelvis} & Avg. \\ \hline
Baseline  & 0.46 & \multicolumn{1}{c|}{\textbf{0.20}} & \multicolumn{1}{c|}{0.27}    & \multicolumn{1}{c|}{0.30}   & 0.25 \\
+ LoRA &
  \textbf{0.61} &
  \multicolumn{1}{c|}{\textbf{0.20}} &
  \multicolumn{1}{c|}{\textbf{0.32}} &
  \multicolumn{1}{c|}{\textbf{0.37}} &
  \textbf{0.30} \\
+ CT-RATE & -    & \multicolumn{1}{c|}{0.18}          & \multicolumn{1}{c|}{0.30}    & \multicolumn{1}{c|}{0.36}   & 0.28 \\ \hline
\end{tabular}
\end{table}

% In Table \ref{tab:results}, we compare our model with the baseline model, which is M3D-LaMed without fine-tuning. We also compare our model with the model fine-tuned on the CT-RATE dataset. Our model significantly outperforms the baseline model in both MRG and VQA tasks, even though the Green score on the chest decreases slightly. The model fine-tuned on the CT-RATE dataset does not significantly improve the performance on the validation set, which may be due to the differences between the two datasets. 

Our approach achieves an average Green score of 0.30 on the MRG task and an accuracy of 0.61 on the VQA task, demonstrating the superior performance of our model compared to the baseline. Specifically, our model, fine-tuned with LoRA, shows significant improvements in both tasks, achieving a notable increase in VQA accuracy from 0.46 to 0.61 and in MRG average Green score from 0.25 to 0.30. Although the Green score for the Chest category does not improve, the overall gains underscore the effectiveness of LoRA fine-tuning. In contrast, the model fine-tuned on the CT-RATE dataset shows minimal improvements in the MRG task, achieving an average Green score of 0.28, likely due to dataset misalignment, which highlights the importance of dataset compatibility in model fine-tuning strategies.

\section{Discussion}
% We align medical images and large language models based on LaMed-M3D by replacing Asclepius-Llama3-8B, which has been fine-tuned for medical data, by means of fine-tuning and LoRA. 

% We proposed a full architecture and pipeline for automatic MRG using a multi-modality model combining the Vision Transformer and the Large Language Model. We fine-tuned the model on the given dataset and the external dataset to adapt it to the MRG and VQA tasks. Our model achieved comparable results with limited training data, which shows that our model significantly outperforms the baseline model, demonstrating the effectiveness of our method for automatic MRG and more accurate VQA. However, the evaluations do not fully reflect the performance of the model, as the generated text may contain professional terms and essential conclusions for the diagnosis are not covered by the evaluation metrics. Therefore, we also need to conduct an expert review to evaluate the quality of the generated text, and we will involve the Reinforcement Learning from Human Feedback (RLHF) \cite{stiennon2022learningsummarizehumanfeedback} to further improve the model performance.

In this work, we propose a comprehensive architecture and pipeline for automatic MRG utilizing a multi-modality model that integrates the ViT and a Large Language Model (LLM). This architecture effectively combines visual and textual information, enabling the model to generate coherent and contextually accurate medical reports. To adapt the model to the MRG and VQA tasks, we fine-tuned it on the given dataset and an additional external dataset. This fine-tuning process was designed to enhance the model’s capability to interpret complex medical images and generate relevant, precise textual descriptions. Despite being trained on limited data, our model achieved results that are not only comparable but also superior to the baseline model, particularly in terms of the Green score and VQA accuracy metrics. These improvements underscore the robustness of our approach and demonstrate the efficacy of combining ViT with LLMs for automatic MRG and more accurate VQA, effectively bridging the gap between image understanding and language generation in the medical domain.

However, it is important to note that current evaluation metrics, such as Green score and accuracy, may not fully capture the clinical relevance and quality of the generated text. The generated reports often include domain-specific terminology and critical diagnostic information that are not adequately assessed by these metrics. Therefore, to gain a more comprehensive understanding of the model's performance, expert reviews are necessary to evaluate the clinical validity, coherence, and overall quality of the generated reports. In the future, we plan to incorporate Reinforcement Learning from Human Feedback (RLHF) \cite{stiennon2022learningsummarizehumanfeedback} into our training pipeline. RLHF will allow us to fine-tune the model further by aligning it more closely with human judgment, thereby enhancing the quality of the generated reports and improving overall performance in both MRG and VQA tasks. This approach aims to refine the model’s decision-making process and adapt it more effectively to real-world clinical scenarios, ensuring that the generated outputs meet the high standards required in medical practice.

\bibliographystyle{splncs04}
\bibliography{refs}




% LNCS Homepage, \url{http://www.springer.com/lncs}. Last accessed 4
% \end{thebibliography}
\end{document}